\documentclass[12pt,preprint]{aastex}

\slugcomment{Submitted to AJ}

\shorttitle{Metal abundances in open clusters}
\shortauthors{Villanova et al.}

\begin{document}


\title{Metal abundances in extremely distant Galactic old open clusters. \\
II. Berkeley 22 and Berkeley 66$^1$}\footnotetext[1]{The data presented
herein were obtained at the W.M. Keck Observatory, which is operated
as a scientific partnership among the California Institute of
Technology, the University of California and the National Aeronautics
and Space Administration. The Observatory was made possible by the
generous financial support of the W.M. Keck Foundation}


\author{Sandro Villanova}
\affil{Dipartimento di Astronomia, Universit\`a di Padova, Vicolo Osservatorio
5, I$-$35122, Padova, Italy}
\email{villanova@pd.astro.it}

\author{Giovanni Carraro\altaffilmark{a,b}}
\affil{Departamento de Astr\'onomia, Universidad de Chile,
    Casilla 36-D, Santiago de Chile, Chile}
\email{gcarraro@das.uchile.cl}

\author{Fabio Bresolin}
\affil{Institute for Astronomy, 2680 Woodlawn Drive, Honolulu, HI 96822, USA}
\email{bresolin@ifa.hawaii.edu}

\and
\author{Ferdinando Patat}
\affil{ESO, K. Schwarzschild Str. 2, 85748 Garching, Germany }
\email{fpatat@eso.org}


\altaffiltext{a}
{Dipartimento di Astronomia, Universit\`a di Padova, Vicolo Osservatorio 5,
I$-$35122, Padova, Italy}
\altaffiltext{b}{Astronomy Department, New Haven, CT 06520$-$8101, USA}


\begin{abstract}
We report on high resolution spectroscopy of four giant stars in
the Galactic old open clusters Berkeley~22 and Berkeley~66
obtained with HIRES at the Keck telescope. We find that
$[Fe/H]=-0.32\pm0.19$ and $[Fe/H]=-0.48\pm0.24$ for Berkeley~22
and Berkeley~66, respectively. Based on these data, we first
revise the fundamental parameters of the clusters, and then
discuss them in the context of the Galactic disk radial abundance
gradient.  We found that both clusters nicely obey the most
updated estimate of the slope of the gradient from \citet{fri02}
and are genuine Galactic disk objects.

\end{abstract}


\section{Introduction}
This paper is the second of a series dedicated at obtaining high
resolution spectroscopy of distant Galactic old open cluster giant
stars to derive new or improved estimates of their metal content.
In \citet{car04} we presented results for Berkeley~29 and
Saurer~1, the two old open clusters possessing the largest
galactocentric distance known, and showed that they do not belong
to the disk, but to the Monoceros feature (see also the discussion in
\citealt{fri05}). Here we present high-resolution spectra of four
giant stars in the old open clusters Berkeley~22 and Berkeley~66.
The latter cluster, in particular, is of particular interest,
since its heliocentric
distance is among the largest currently known.\\
With this paper we aim to enlarge the sample of old open clusters
with metallicity obtained from high resolution spectra, and to
test the common assumption of axisymmetry made in chemical
evolution models about the structure of the Milky Way disk. For
this purpose we selected two open clusters with about the same age
located one (Berkeley~66) in the second Galactic Quadrant, and the
other (Berkeley~22) in the third Galactic Quadrant. Significant
differences in  metal abundance for clusters located in different
disk zones would hopefully provide useful clues about the chemical
evolution
of the disk and about the role of accretion and infall phenomena.\\
\noindent The layout of the paper is as follows. Sections~2 and 3
illustrate the observations and the data reduction strategies,
while Section~4 deals with radial velocity determinations. In
Section~5 we derive the stellar abundances and in Section~6 we
revise the cluster fundamental parameters. 
The radial abundance gradient and the abundance ratios are discussed 
in Section 7 and 8, respectively.
The results of this
paper are finally  discussed in Section~9

\section{Observations}
The observations were carried out on the night of November 30,
2004 at the W.M. Keck Observatory under photometric conditions and
typical seeing of 1\farcs1 arcsec.  The HIRES spectrograph
\citep{vog94} on the Keck I telescope was used with a 1\farcs1 x
7\arcsec\/ slit to provide a spectral resolution R = 34,000 in the
wavelength range 5200$-$8900~\AA\/ on the three 2048$\times$4096
CCDs of the mosaic detector. A blocking filter was used to remove
second-order contamination from blue wavelengths. Three exposures
of 1500--1800 seconds were obtained for the stars
Berkeley~22$-$400 and Berkeley~22$-$579. For both
Berkeley~66$-$785 and Berkeley~66$-$934 we took four exposures of
2700 seconds each. During the first part of the night, when the
telescope pointed to Berkeley~66, the observing conditions were
far from optimal, due to the presence of thick clouds. For
this reason our abundance analysis in this first cluster is
limited to Berkeley~66$-$785. For the wavelength calibration,
spectra of a thorium-argon lamp were secured after the set of
exposures for each star was completed. The radial
velocity standard HD~82106 was observed at the end of the night.\\
In Fig.~1 we show a finding chart for the two clusters where the
four observed stars are indicated, while in Fig.~2 we show the
position of the stars in the Color$-$Magnitude Diagram (CMD),
based on published photometry (\citealt{kal94}, \citealt{phe96}).

\section{Data Reduction}
Images were reduced using IRAF\footnote[2]{IRAF is distributed by
the National Optical Astronomy Observatories, which are operated
by the Association of Universities for Research in Astronomy,
Inc., under cooperative agreement with the National Science
Foundation.}, including bias subtraction, flat-field correction,
frame combination, extraction of spectral orders, wavelength
calibration, sky subtraction and spectral rectification. The
single orders were merged into a single spectrum. As an example,
we show in Fig.~3 a portion of the reduced, normalized spectrum of
Be~22-400.

\section{Radial Velocities}
No radial velocity estimates were previously available for
Berkeley~22 and Berkeley~66. The radial velocities of the target
stars were measured using the IRAF FXCOR task, which
cross-correlates the object spectrum with the template (HD~82106).
The peak of the cross-correlation was fitted with a Gaussian curve
after rejecting the spectral regions contaminated by telluric
lines ($\lambda > 6850$~\AA).  In order to check our wavelength
calibration we also measured the radial velocity of HD~82106
itself, by cross-correlation with a solar-spectrum template. We
obtained a radial heliocentric velocity of 29.8$\pm$0.1 km
s$^{-1}$, which perfectly matches the published value (29.7
km/sec; \citealt{Udr99}). The final error in the radial velocities
was typically about 0.2 km s$^{-1}$.  The two stars we measured in
each clusters have compatible radial velocities (see Table~1), and
are considered, therefore,  {\it bona fide} cluster members.

\section{ABUNDANCE ANALYSIS}

\subsection{Atomic parameters and equivalent widths}
We derived equivalent widths of spectral lines by using the
standard IRAF routine {\it SPLOT}. Repeated measurements show a
typical error of about 5$-$10 m\AA, also for the weakest lines.
The line list (FeI, FeII, Mg, Si, Ca, Al,, Na, Ni and Ti, see
Table~3) was taken from \citet{fri03},
who considered only lines with equivalent widths narrower than
150m\AA, in order to avoid non-linear effects in the LTE analysis
of  the spectral features. log(gf) parameters of these lines were
re-determinated using equivalent widths from the solar-spectrum
template, solar abundances from \citet{and89} and standard
solar parameters ($T_{eff}=5777~K, log(g)=4.44, v_t=0.8~Km\,
s^{-1}$).

\subsection{Atmospheric parameters}
Initial estimates of the atmospheric parameter $T_{eff}$ were
obtained from photometric observations in the optical.  BVI data
were available for Berkeley~22 \citep{kal94}, while VI photometry
for Berkeley~66 has been taken from \citep{phe96}.  Reddening
values are E(B$-$V)= 0.62 (E(V$-$I)= 0.74), and E(V$-$I)= 1.60,
respectively. First guess effective
temperatures were derived from the (V$-$I)--$T_{eff}$ and
(B$-$V)--$T_{eff}$ relations, the former from \citet{alo99} and the
latter from \citet{gra96}.  We then adjusted the effective
temperature by minimizing the slope of the abundances obtained
from Fe~I lines with respect to the excitation potential in the
curve of growth analysis. For both clusters the derived
temperature
yields a reddening consistent with the photometric one.\\
\noindent
Initial guesses for the gravity $\log$(g) were derived from:

\begin{equation}
log(\frac{g}{g_{\odot}}) = log(\frac{M}{M_{\odot}}) + 4 \times
log(\frac{T_{eff}}{T_{\odot}}) - log(\frac{L}{L_{\odot}})
\end{equation}

\noindent
taken from \citet{carre97}.  In this equation the mass
$\frac{M}{M_{\odot}}$ was derived from the comparison between the
position of the star in the Hertzsprung$-$Russell diagram and the Padova
Isochrones \citep{gir00}. The luminosity $\frac{L}{L_{\odot}}$
was derived from the the absolute magnitude $M_V$, assuming
the literature distance moduli of
15.9 for Berkeley~22 \citep{kal94} and 17.4 for Berkeley~66 \citep{phe96}.
The bolometric
correction (BC) was derived from the relation BC--$T_{eff}$ from
\citet{alo99}.  The input $\log$(g) values were then adjusted in order to
satisfy the ionization equilibrium of Fe~I and Fe~II during the
abundance analysis.  Finally, the micro$-$turbulence velocity is given by
the following relation \citep{gra96}:

\begin{equation}
v_t[km\,s^{-1}] = 1.19 \times 10^{-3} \times T_{eff} - 0.90 \times
log(g) - 2
\end{equation}

\noindent
The final adopted parameters are listed
in Table~2.

\subsection{Abundance determination}
The LTE abundance program MOOG (freely distributed by Chris
Sneden, University of Texas, Austin) was used to determine the
metal abundances. Model atmospheres were interpolated from the
grid of Kurucz (1992) models  by using the values of $T_{eff}$ and
$\log$(g) determined as explained in the previous section. During
the abundance analysis $T_{eff}$, $\log$(g) and $v_t$ were
adjusted to remove trends in excitation potential, ionization
equilibrium and equivalent width for Fe~I and Fe~II lines. Table~3
contains the atomic parameters and equivalent widths for the lines
used. The first column contains the name of the element, the
second the wavelength in \AA, the third the excitation potential,
the fourth the oscillator strength $\log$ ({\it gf}), and the
remaining ones the equivalent widths of the lines for the observed
stars. \\The derived abundances are listed in Table~4, together
with their uncertainties. The measured iron abundances are
[Fe/H]=$-$0.32$\pm0.19$ and [Fe/H]=$-$0.48$\pm0.24$ for
Berkeley~22 and Berkeley~66, respectively. The reported errors are
derived from the uncertainties on the single star abundance
determination (see Table~4). For Berkeley~66$-$934 no abundance
determination was possible because the S/N ratio in our spectrum
was too low to perform any equivalent width determination.\\

\noindent Finally, using the stellar parameters (colors, T$_{eff}$
and log(g)) and the absolute calibration of the MK system
(\citealt{str81}), we derived the stellar spectral classification,
which we provide in Table~1.

\section{REVISION OF CLUSTER PROPERTIES}
Our study is the first to provide spectral abundance determinations of
stars in Berkeley~22 and Berkeley~66. Here we briefly discuss the
revision of the properties of these two clusters which follow from our
measured chemical abundances (see Figs.~4 and 5). We use the isochrone
fitting method and adopt for this purpose the Padova models from
\cite{gir00}.

\subsection{Berkeley 22}
Berkeley~22 is an old open cluster located in the third Galactic
quadrant, first studied by \citet{kal94}. On the basis of deep VI
photometry he derived an age of 3 Gyr, a distance of 6.0 Kpc and a
reddening E(V$-$I)=0.74. The author suggests that the probable
metal content of the cluster is lower than solar. Here we obtained
[Fe/H]=$-$0.32, which corresponds to Z=0.008 and roughly
half the solar metal content. 
Very recently \citet{dif05} presented new BVI photometry, on the 
basis of which they derive a younger age (2.0-2.5 Gyr), 
but similar reddening and distance, also suggesting that
the cluster posseses solar metal abundance.
The two photometric studies are compatible in the VI filters, 
being the difference in the V and I zeropoints 
less than 0.03 mag. Both the studies show that the cluster
turnoff is located at V = 19 and a clump at V = 16.65.\\
According to \citet{car94}, with a $\Delta V \approx 2.35$ mag
($\Delta V$ being the magnitude difference between the turnoff and the clump)
and for the derived metallicity, one would expect an age around 3.5
Gyr. This preliminary
estimate of the age is in fact confirmed by the isochrone fitting method.
In Fig.~4   
the solid line is a 3.3 Gyr isochrone, which provides 
a new age estimate of 3.3$\pm$0.3 Gyrs for the same
photometric reddening and Galactocentric distance derived by
\cite{kal94}. Both the turnoff and the clump location
are nicely reproduced. The uncertainty reflects the range
of isochrones which produce an acceptable fit.\\
For comparison, in the same figure 
we over-impose a solar metallicity isochrone
for the age of 2.25 Gyr (dashed line in Fig.~4),
and for the same reddening and distance reported by  \citet{dif05}.
This isochrone clearly does not provide a
comparable good fit. When trying to fit the turnoff,
both the color of the Red Giant Branch  and the position
of the clump cannot be reproduced.

\subsection{Berkeley 66}
Berkeley~66 was studied by \citet{gua97} and \citet{phe96}, who
suggested a reddening E(V$-$I)=1.60 and a  metallicity in the
range $-0.23 \leq [Fe/H] \leq 0.0$. By assuming these values,
\citet{phe96} derived a galactocentric distance of 12.9 Kpc and an
age of 3.5 Gyrs. We obtain a significantly smaller abundance value
[Fe/H]=$-$0.48$\pm$0.24 for a spectroscopic reddening of
E(V$-$I)=1.60, which is consistent with the \citet{phe96}
estimate. 
By looking at the CMD, the turnoff is situated at V = 20.75,
whereas  the clump is at V =  18.25, implying a $\Delta V$
of 2.5. For this  $\Delta V$ and the derived [Fe/H],
the  \citet{car94} calibration yields an age of 
about 4.7 Gry, significantly larger than previous estimate.
The value  [Fe/H]=$-$0.48 tanslates into Z=0.006,
and we generated a few isochrones for this exact
metal abundance from \cite{gir00}. 
The result is shown in Fig~5, where
a 4 and 5 Gyr isochrones (dashed and solid line, respectively)
are overimposed to the cluster CMD.
Both the isochrones reasonably reproduce the turnoff shape but the 5 Gyr
one (solid line) clearly presents a too bright clump.
On the other hand, the 4 Gyr isochrone reproduces well all
the CMD feature. This new age estimate provides a reddening E(V-I) = 1.60
and a heliocentric distance of  5 Kpc.

\section{THE RADIAL ABUNDANCE GRADIENT}
In Fig.~6 we plot the open cluster Galactic radial abundance
gradient, as derived from \citet{fri02}, which is at present  the
most updated version of the gradient itself. The clusters included
in their work (open squares) define an overall slope of
$-$0.06$\pm$0.01 dex Kpc$^{-1}$ (solid line). The filled circles
represent the two clusters analyzed here, and it can be seen that
they clearly follow very nicely the general trend. In fact the
dashed line, which represents the radial abundance gradient
determined by including Berkeley~22 and Berkeley~66, basically
coincides with the \citet{fri02} one.\\
We note that the gradient
exhibits quite a significant scatter.  One may wonder whether this
solely depends on observational errors, or whether this scatter
reflects a true chemical inhomogeneity in the Galactic disk.\\
\noindent
It must be noted however that between 10 and 14 kpc the scatter increases
and the distribution of the cluster abundances is compatible also
with a flat gradient. Again, it is extremely difficult to conclude whether
this behaviour of the gradient is significant, or whether
the distribution of abundances is mostly affected by the size
of the observational errors and the small number of clusters involved.

\section{ABUNDANCE RATIOS}
Abundance ratios constitute a powerfull tool to assign a cluster to a 
stellar population \citealt{fri03}.
In PaperI (\citealt{car04}) we found that Saurer~1 and Berkeley~29 
exhibit enhanced abundance ratio with respect to the Sun, and we concluded
 that they probably do not belong to the Galactic disk, since
all the old open cluster for which detailed abundance analysis is available
show solar scaled abundance ratios.\\
In Table~5 we list the abundance ratios for the observed stars in
Berkeley~22 and Berkeley~66. Our program clusters have ages around
3-4 Gyrs and iron metal content [Fe/H] $\approx$ $-$0.3-0.4. They
are therefore easily comparable with similar clusters from the
literature, such as Tombaugh~2 and Melotte~66 (see
\citealt{fri03}, Tab.~7). We note that the
latter two clusters and our program clusters have  scaled solar
abundances. Indeed, at a similar [Fe/H] and age , Berkeley~22 and
Melotte~66 have  similar values for
all the abundance ratios.\\
Similar conclusions can be drawn for Berkeley~66, when compared
with Tombaugh~2, an old open cluster of similar age and [Fe/H].
All the abundance ratios we could measure are comparable in these two clusters.\\
\noindent
These results therefore indicate that Berkeley~22 and Berkeley~66
are two genuine old disk clusters, which well fit in the overall
Galactic radial abundance gradient (see previous section).

\section{DISCUSSION and CONCLUSIONS}
Berkeley~22 and Berkeley~66 are two similar age open clusters
located roughly at symmetric positions with respect to the virtual
line connecting the Sun to the Galactic center. In fact for
Berkeley~22 we derive +5.5, $-$2.0 and $-$0.8 Kpc for the
rectangular Galactic coordinate X, Y and Z, respectively, while
for Berkeley~66 we obtain +4, +2.5 and 0.01 Kpc. The corresponding
Galactocentric distances $R_{GC}$
are 12.7 and 14.2 Kpc for Berkeley~66 and Berkeley~22, respectively.\\
Within the errors the two clusters possess the same metal
abundance, suggesting that at the distance of 12-14 Kpc from the
Galactic Center
the metal distribution over the second and third quadrant of the Galaxy is 
basically the same.\\
It is worthwhile to point out here that only 3 clusters are insofar 
known to lie outside the 14 Kpc-radius ring from the Galactic center:
Berkeley~20, Berkeley~29 and Saurer~1. \citet{car04} showed that
both Berkeley~29 and Saurer~1 do not belong to the disk, and
therefore we are probably sampling here the real outskirts of the
Galactic stellar disk.\\

\noindent 
The axisymmetric homogeneity
is confirmed when we add a few more old open
clusters located in this strip, like NGC 1193, NGC~2158, NGC~2141,
Berkeley~31, Tombaugh~2  and Berkeley~21 (\citealt{fri02}). All
these clusters are located in the second and third quadrant, have
metallicities $ -0.62 \leq [Fe/H] \leq -0.25$ and probably belong all to
the same generation (ages between 2 and 4 Gyr). 
Therefore, although with a significant
spread, we conclude that at 12-14 Kpc the disk is chemically
homogeneous in [Fe/H] within the observational errors. The
scatter, if real, may be due to local inhomogeneities.

\acknowledgments
The work of GC is supported by {\it Fundacion Andes}.
We thank the anonymous referee for his/her report
which helped a lot to improve on the paper presentation.

\clearpage



\clearpage

\begin{figure}
\plottwo{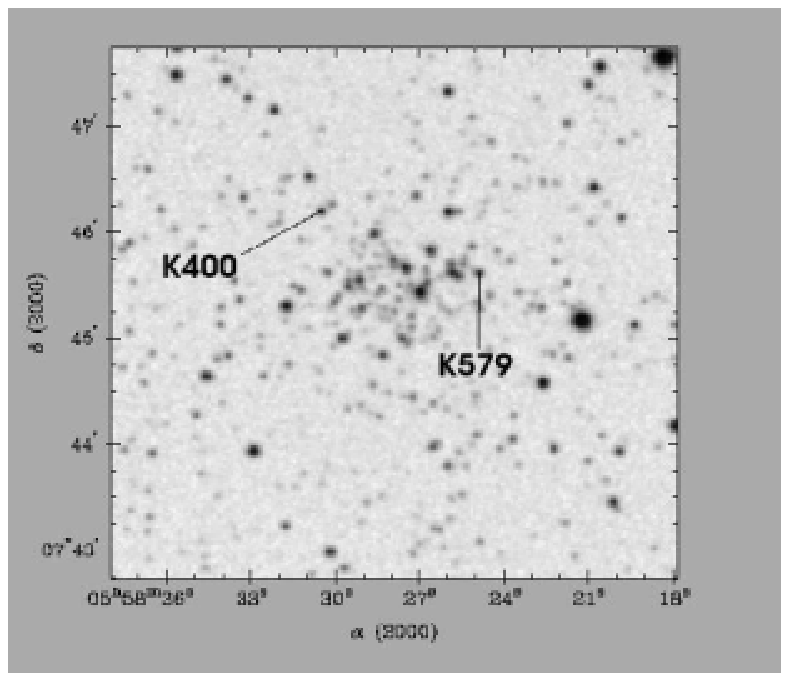}{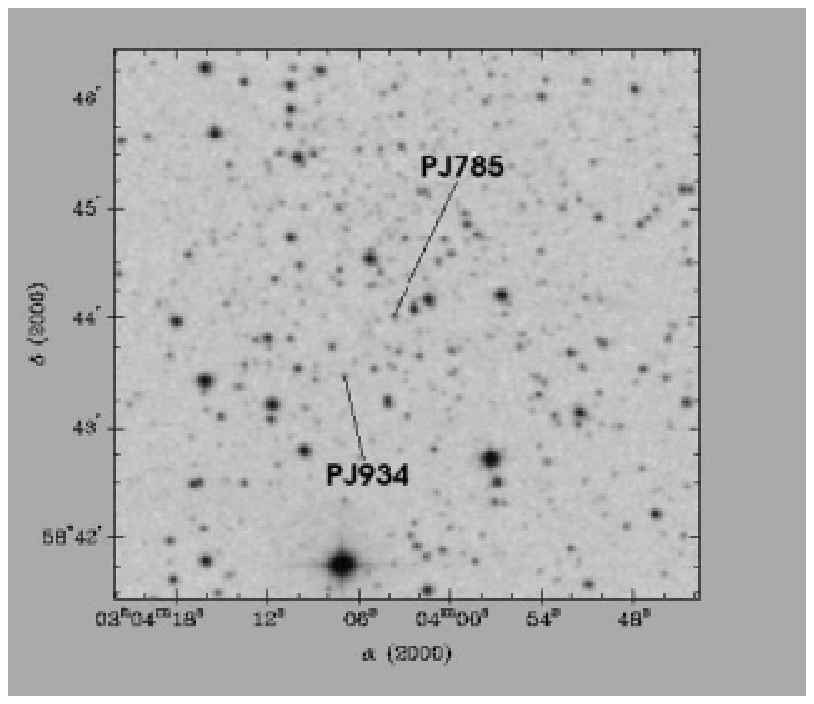} \caption{Digital
Sky Survey finding charts of the observed stars in Berkeley~22
(left panel) and Berkeley~66 (right panel)}
\end{figure}

\begin{figure}
\includegraphics[scale=0.8]{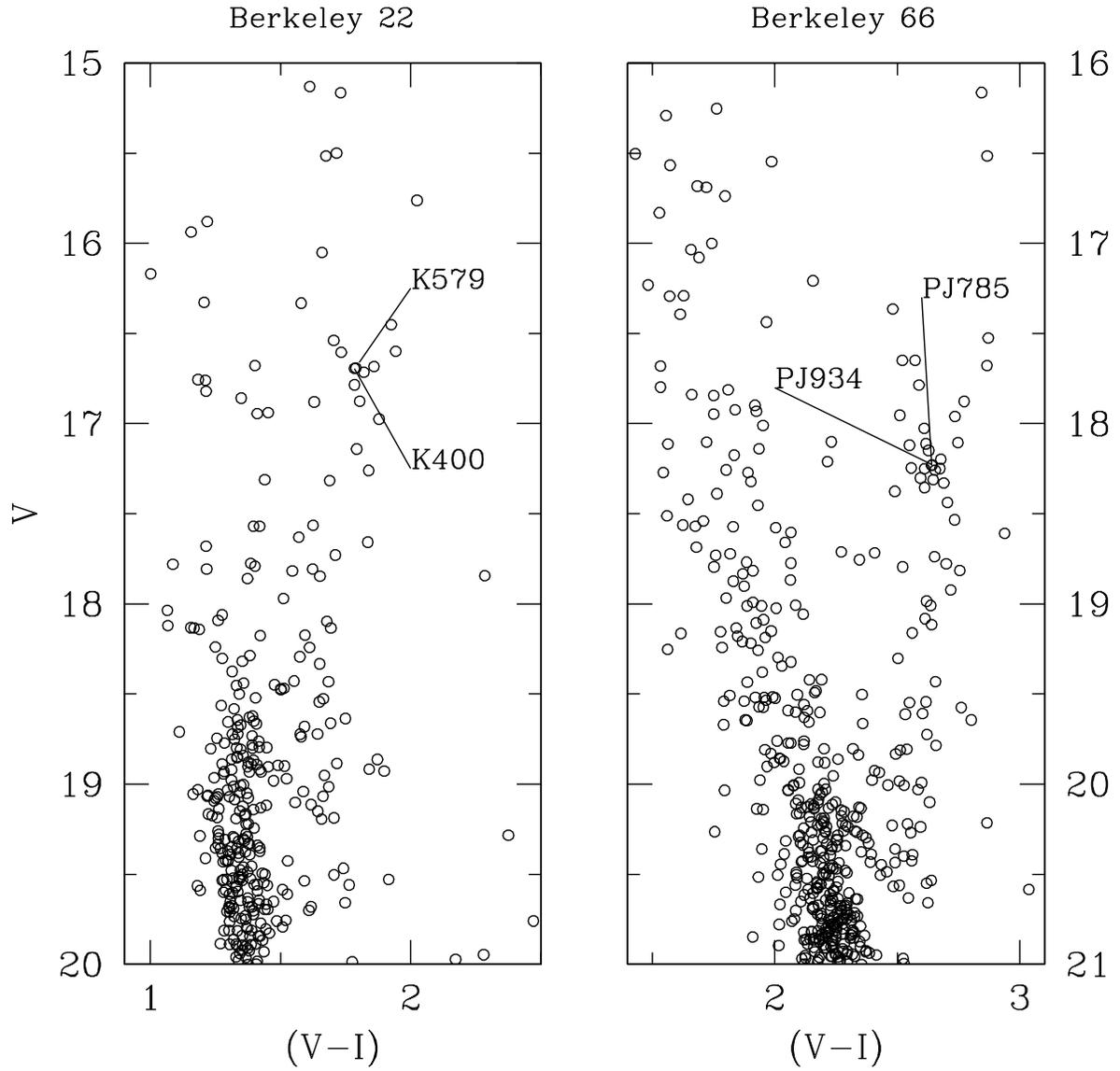}
\caption{Position of the observed stars in the CMD of Berkeley~22 (left panel, photometry from  \citealt{kal94})
and Berkeley~66 (right panel, photometry from \citealt{phe96}).}
\end{figure}

\begin{figure}
\includegraphics[scale=0.8]{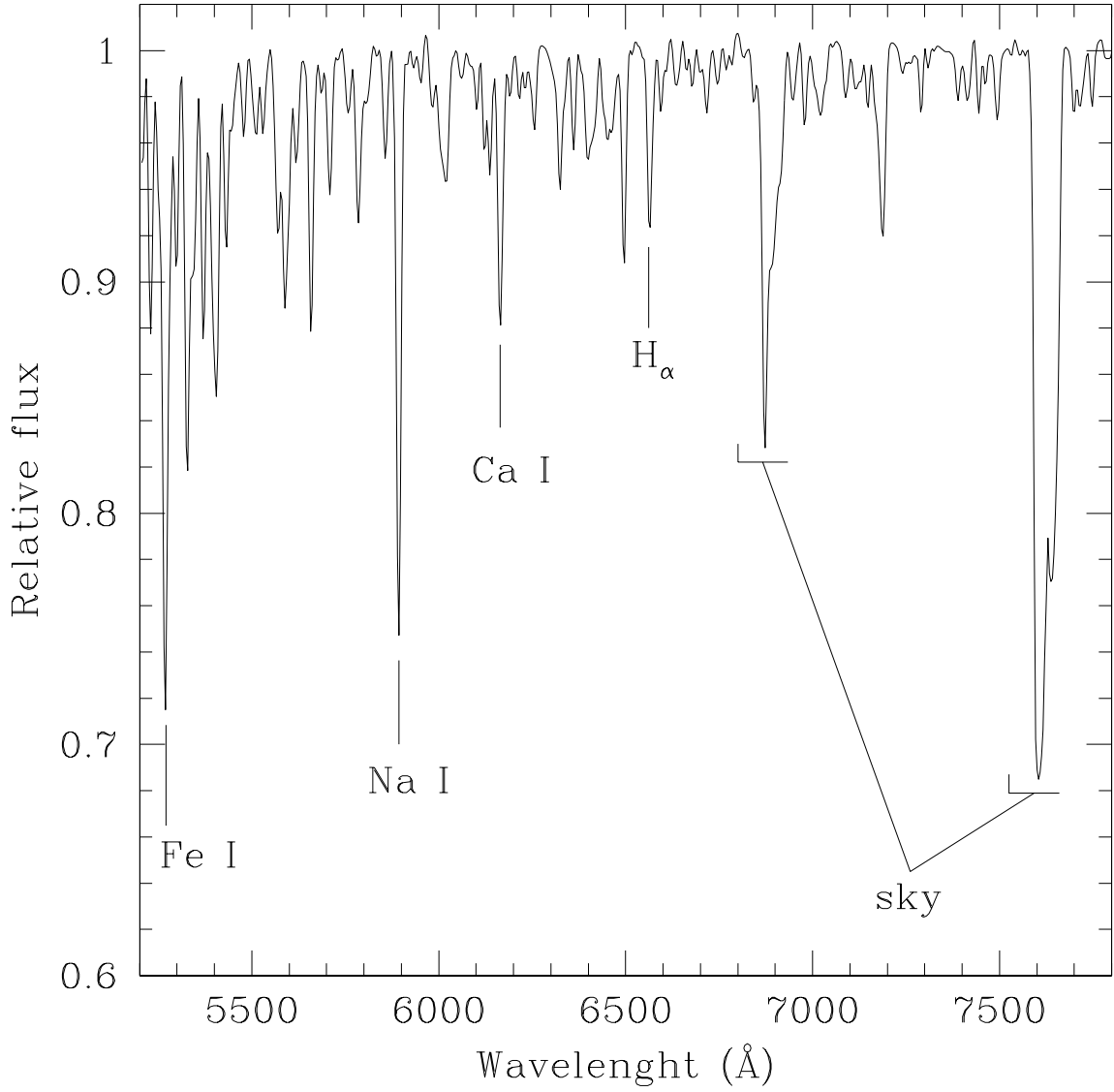}
\caption{An example of extracted spectrum for the star Be~22-400,
with the main lines indicated}
\end{figure}

\begin{figure}
\includegraphics[scale=0.8]{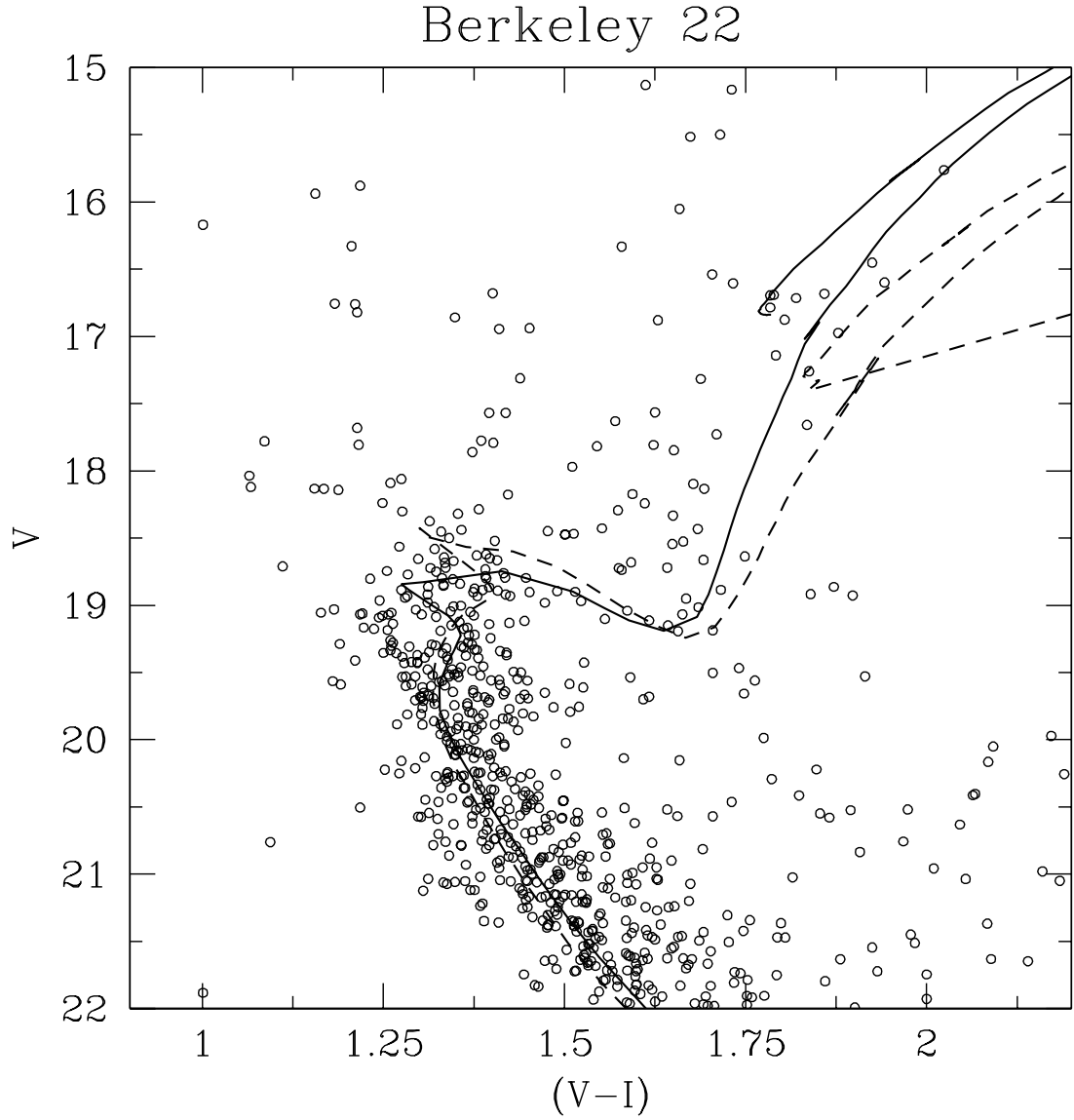}
\caption{Isochrone solution for Berkeley~22 (photometry from  \citealt{kal94})
The solid line is a 3.3 Gyr isochrone for Z=0.008, whilst
the dashed one is a 2.25 Gyr isochrone for Z=0.019. 
See text for details.}
\end{figure}

\begin{figure}
\includegraphics[scale=0.8]{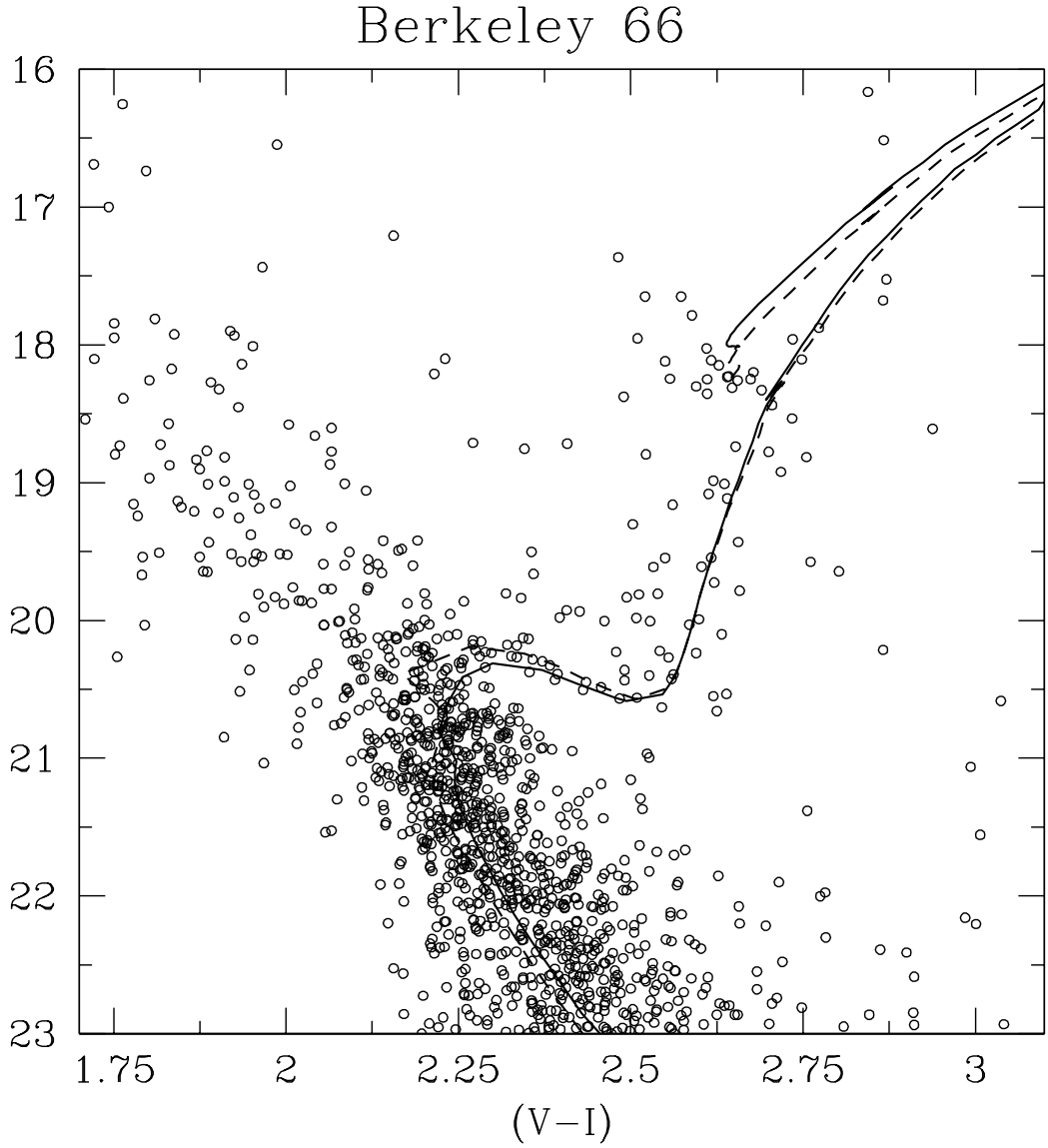}
\caption{Isochrone solution for Berkeley~66 (photometry from \citealt{phe96}). 
The solid and dashed lines are two Z=0.008
isochrones, for the age of 5 and 4 Gyrs, 
respectively.
See text for details.}
\end{figure}

\begin{figure}
\includegraphics[scale=0.8]{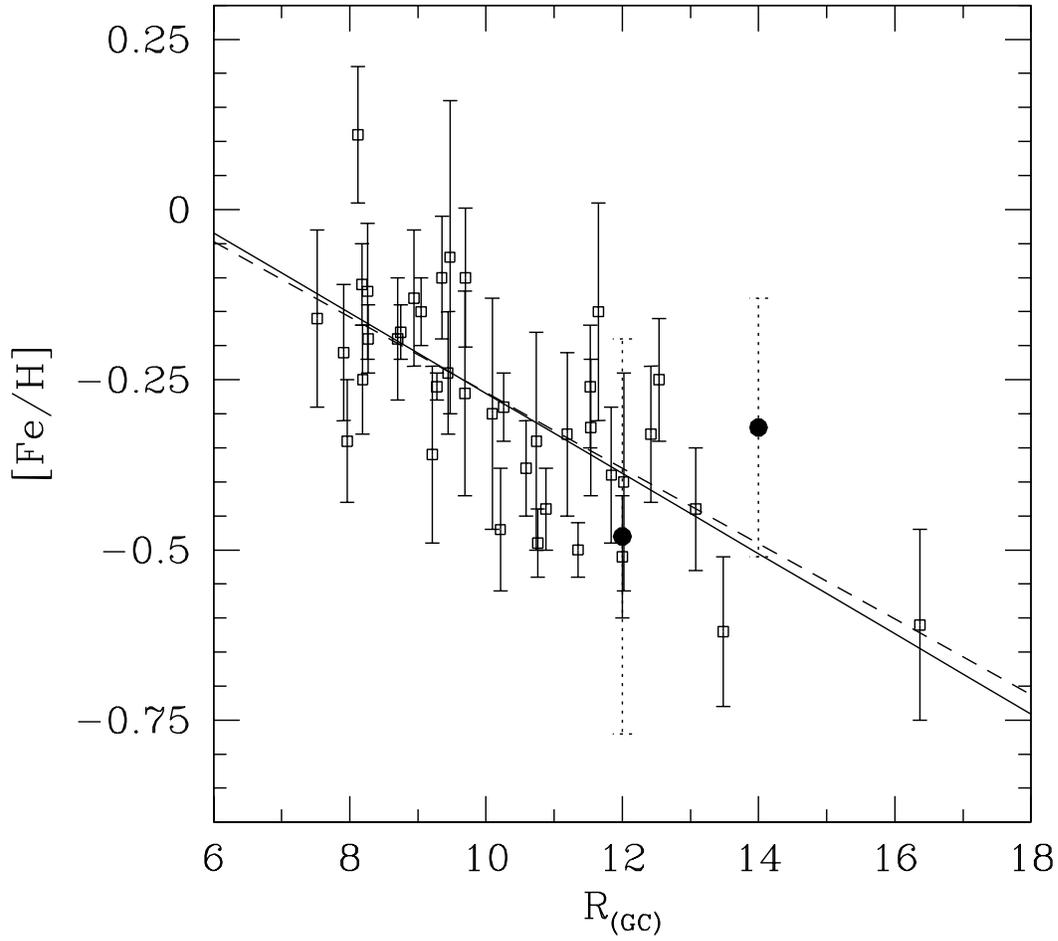}
\caption{The Galactic disk chemical abundance radial gradient.
Open squares are data from \citet{fri02}, whereas filled circles
represent Berkeley~22 and Berkeley~66 (this work). The solid line
is the linear fit to the \citet{fri02} data, whereas the dashed
line is a linear fit to all the data points.}
\end{figure}

\clearpage

\begin{deluxetable}{lccccccccc}
\tabletypesize{\scriptsize}
\tablewidth{0pt}
\tablecaption{OBSERVED STARS}
\tablehead{
\colhead{ID} & \colhead{RA} & \colhead{DEC} & \colhead{V} & \colhead{(B-V)}  & \colhead{(V$-$I)} & \colhead{$V_{rad}$ (km s$^{-1}$)} & \colhead{$S/N$} & \colhead{Spectral Type} & \colhead {comments}}
\startdata
Be 22-400  & 05:58:30.97 & +07:46:15.3 & 16.70 &  1.58   & 1.78 &  93.3$\pm$0.2 & 25 & G8III  & \citet{kal94}\\
Be 22-579  & 05:58:25.78 & +07:45:31.2 & 16.88 &  1.66   & 1.80 &  97.3$\pm$0.2 & 20 & K0III  & \citet{kal94}\\
Be 66-785  & 03:04:02.90 & +58:43:57.0 & 18.23 &\nodata  & 2.64 & -50.7$\pm$0.1 & 15 & K1III  & \citet{phe96}\\
Be 66-934  & 03:04:06.41 & +58:43:31.0 & 18.23 &\nodata  & 2.64 & -50.6$\pm$0.3 &  5 & K1III  & \citet{phe96}\\
\enddata
\end{deluxetable}

\begin{deluxetable}{lccc}
\tabletypesize{\scriptsize}
\tablewidth{0pt}
\tablecaption{ADOPTED ATMOSPHERIC PARAMETERS}
\tablehead{
\colhead{ID} & \colhead{T$_{eff(K)}$} & \colhead{log {\it g} (dex)} & \colhead{$v_t$ (km s$^{-1}$)} }
\startdata
Be 22-400  & 4790$\pm$100 & 2.8$\pm$0.1 & 1.3 \\
Be 22-579  & 4690$\pm$50  & 2.8$\pm$0.1 & 1.2 \\
Be 66-785  & 4640$\pm$100 & 2.7$\pm$0.1 & 1.2 \\
Be 66-934  &  \nodata     &  \nodata    & \nodata  \\
\enddata
\end{deluxetable}

\clearpage

\begin{deluxetable}{lcccccc}
\tabletypesize{\scriptsize}
\tablewidth{0pt}
\tablecaption{EQUIVALENT WIDTHS}
\tablehead{
\colhead{Element} & \colhead{$\lambda (\AA)$} & \colhead{$E.P.$} & \colhead{$log\;{\it gf}$} & \colhead{$Be22-400$} & \colhead{$Be22-579$} &  \colhead{$Be66-785$}}
\startdata
 Fe I & 5379.570 & 3.680   & -1.57 &    83  &   102 &\nodata\\
 Fe I & 5417.033 & 4.415   & -1.45 &    45  &    44 &    54 \\
 Fe I & 5466.988 & 3.573   & -2.24 &    69  &\nodata&\nodata\\
 Fe I & 5633.946 & 4.990   & -0.23 &    68  &    86 &\nodata\\
 Fe I & 5662.520 & 4.160   & -0.59 &   106  &   111 &\nodata\\
 Fe I & 5701.550 & 2.560   & -2.19 &   125  &   113 &   146 \\
 Fe I & 5753.120 & 4.240   & -0.76 &   102  &   108 &    94 \\
 Fe I & 5775.081 & 4.220   & -1.17 &    96  &    77 &    82 \\
 Fe I & 5809.218 & 3.883   & -1.67 &    80  &    83 &\nodata\\
 Fe I & 6024.058 & 4.548   & +0.03 &   124  &   108 &   104 \\
 Fe I & 6034.036 & 4.310   & -2.35 &    58  &\nodata&\nodata\\
 Fe I & 6056.005 & 4.733   & -0.48 &    84  &    82 &    71 \\
 Fe I & 6082.72  & 2.22    & -3.62 &    77  &    76 &\nodata\\
 Fe I & 6093.645 & 4.607   & -1.38 &    47  &\nodata&\nodata\\
 Fe I & 6096.666 & 3.984   & -1.82 &    47  &    46 &\nodata\\
 Fe I & 6151.620 & 2.180   & -3.34 &    63  &   103 &    81 \\
 Fe I & 6165.360 & 4.143   & -1.53 &    47  &    66 &\nodata\\
 Fe I & 6173.340 & 2.220   & -2.90 &   115  &   101 &   104 \\
 Fe I & 6200.313 & 2.608   & -2.38 &   124  &   102 &    81 \\
 Fe I & 6229.230 & 2.845   & -2.93 &    82  &\nodata&\nodata\\
 Fe I & 6246.320 & 3.590   & -0.77 &   119  &\nodata&\nodata\\
 Fe I & 6344.15  & 2.43    & -2.90 &   116  &   107 &    78 \\
 Fe I & 6481.880 & 2.280   & -2.95 &   121  &   109 &    97 \\
 Fe I & 6574.229 & 0.990   & -5.11 &    97  &    92 &    90 \\
 Fe I & 6609.120 & 2.560   & -2.67 &   108  &    93 &   125 \\
 Fe I & 6703.570 & 2.758   & -3.08 &    78  &    71 &    52 \\
 Fe I & 6705.103 & 4.607   & -1.07 &    57  &    60 &\nodata\\
 Fe I & 6733.151 & 4.638   & -1.48 &\nodata &\nodata&    37 \\
 Fe I & 6810.263 & 4.607   & -0.99 &    64  &    72 &\nodata\\
 Fe I & 6820.372 & 4.638   & -1.14 &    50  &    56 &\nodata\\
 Fe I & 6839.831 & 2.559   & -3.42 &    70  &    67 &    74 \\
 Fe I & 7540.430 & 2.730   & -3.87 &    56  &\nodata&\nodata\\
 Fe I & 7568.900 & 4.280   & -0.85 &    80  &    92 &    91 \\
 Fe II& 5414.080 & 3.22    & -3.60 &    27  &\nodata&\nodata\\
 Fe II& 6084.100 & 3.20    & -3.78 &    21  &\nodata&\nodata\\
 Fe II& 6149.250 & 3.89    & -2.67 &    36  &\nodata&\nodata\\
 Fe II& 6247.560 & 3.89    & -2.31 &    63  &    48 &\nodata\\
 Fe II& 6369.463 & 2.89    & -4.18 &    28  &\nodata&\nodata\\
 Fe II& 6456.390 & 3.90    & -2.05 &    50  &    57 &\nodata\\
 Fe II& 6516.080 & 2.89    & -3.24 &\nodata &    56 &\nodata\\
 Al I & 6696.03  & 3.14    & -1.56 &     71 &\nodata&    43 \\
 Al I & 6698.67  & 3.13    &       &\nodata &    44 &\nodata\\
 Ca I & 5581.97  & 2.52    & -0.62 &    114 &   119 &   107 \\
 Ca I & 5590.12  & 2.52    & -0.82 &    105 &   112 &   100 \\
 Ca I & 5867.57  & 2.93    & -1.65 &     45 &\nodata&\nodata\\
 Ca I & 6161.30  & 2.52    & -1.27 &     93 &   109 &    92 \\
 Ca I & 6166.44  & 2.52    & -1.12 &     94 &    83 &   103 \\
 Ca I & 6455.60  & 2.52    & -1.41 &     82 &    70 &    72 \\
 Ca I & 6499.65  & 2.52    & -0.91 &    117 &   111 &    85 \\
 Mg I & 5711.09  & 4.33    & -1.71 &    117 &   109 &\nodata\\
 Mg I & 7387.70  & 5.75    & -1.09 &    105 &    67 &\nodata\\
 Na I & 5682.65  & 2.10    & -0.75 &    126 &   128 &\nodata\\
 Na I & 5688.21  & 2.10    & -0.72 &    136 &   138 &   160 \\
 Na I & 6154.23  & 2.10    & -1.61 &     56 &    45 &    51 \\
 Na I & 6160.75  & 2.10    & -1.38 &     73 &    73 &    64 \\
 Ni I & 6175.37  & 4.09    & -0.52 &     74 &    64 &    62 \\
 Ni I & 6176.81  & 4.09    & -0.19 &    116 &    78 &    80 \\
 Ni I & 6177.25  & 1.83    & -3.60 &     36 &    42 &\nodata\\
 Ni I & 6223.99  & 4.10    &       &\nodata &\nodata&    64 \\
 Si I & 5665.60  & 4.90    & -1.98 &\nodata &    53 &\nodata\\
 Si I & 5684.52  & 4.93    & -1.63 &     56 &\nodata&\nodata\\
 Si I & 5701.12  & 4.93    & -1.99 &\nodata &    41 &\nodata\\
 Si I & 5793.08  & 4.93    & -1.89 &     38 &    49 &\nodata\\
 Si I & 6142.49  & 5.62    & -1.47 &     27 &\nodata&\nodata\\
 Si I & 6145.02  & 5.61    &       &\nodata &\nodata&\nodata\\
 Si I & 6243.82  & 5.61    & -1.30 &     39 &    54 &\nodata\\
 Si I & 7034.91  & 5.87    & -0.74 &     55 &   103 &\nodata\\
 Ti I & 5978.54  & 1.87    & -0.65 &     71 &    67 &    67 \\
 Ti II& 5418.77  & 1.58    & -2.12 &     74 &    76 &   103 \\
\enddata
\end{deluxetable}

\begin{deluxetable}{lccccccccc}
\tabletypesize{\scriptsize}
\tablewidth{0pt}
\tablecaption{MEAN STELLAR ABUNDANCES}
\tablehead{
\colhead{ID} & \colhead{[FeI/H]} &  \colhead{[FeII/H]} & \colhead{[AlI/H]} & \colhead{[CaI/H]} & \colhead{[MgI/H]} &  \colhead{[NaI/H]} & \colhead{[NiI/H]} & \colhead{[SiI/H]} & \colhead{[Ti/H]} }
\startdata
Be 22-400  & -0.29$\pm$0.21 & -0.32$\pm$0.19 & +0.05$\pm$0.20 & -0.35$\pm$0.11 & -0.37$\pm$0.20 & -0.23$\pm$0.05 & -0.26$\pm$0.20 & -0.37$\pm$0.08 & -0.19$\pm$0.11\\
Be 22-579  & -0.35$\pm$0.17 & -0.28$\pm$0.04 & -0.12$\pm$0.20 & -0.45$\pm$0.11 & -0.41$\pm$0.14 & -0.33$\pm$0.12 & -0.29$\pm$0.07 & -0.20$\pm$0.11 & -0.22$\pm$0.04\\
Be 66-785  & -0.48$\pm$0.24 &    \nodata     & -0.48$\pm$0.20 & -0.53$\pm$0.21 &   \nodata      & -0.33$\pm$0.18 & -0.24$\pm$0.25 &    \nodata     & -0.05$\pm$0.23\\
\enddata
\end{deluxetable}

\begin{deluxetable}{lcccccccc}
\tabletypesize{\scriptsize}
\tablewidth{0pt}
\rotate
\tablecaption{ABUNDANCE RATIOS}
\tablehead{
\colhead{ID} & \colhead{[Fe/H]} &  \colhead{[Ca/Fe]} & \colhead{[Mg/Fe]} & \colhead{[Si/Fe]} & \colhead{[Ti/Fe]} & \colhead{[Na/Fe]} & \colhead{[Al/Fe]} & \colhead{[Ni/Fe]} }
\startdata
Be 22-400  & -0.29  & -0.06 &  -0.08  &  -0.08  & +0.10 & +0.06 & +0.34 & +0.03\\
Be 22-579  & -0.35  & -0.10 &  -0.06  &  +0.15  & +0.13 & +0.02 & +0.23 & +0.06\\
Be 66-785  & -0.48  & -0.05 & \nodata & \nodata & +0.43 & +0.15 & +0.00 & +0.24\\
\enddata
\end{deluxetable}



\clearpage


\begin{thebibliography}{}

\bibitem[Alonso et al.\/(1999)]{alo99} Alonso A., Arribas S.,
Mart\'inez-Roger C. 1999, \aap~ 140, 261

\bibitem[Anders \& Grevesse\/(1989)]{and89} Anders E., Grevesse N. 1989,
GeCoA 53, 197

\bibitem[Carraro \& Chiosi\/(1994)]{car94} Carraro G. \& Chiosi C.
1994, \aap~ 287, 761

\bibitem[Carraro et al.\/(2004)]{car04} Carraro G., Bresolin F., Villanova S., Matteucci F.,
Patat F., Romaniello M. 2004, \aj~ 128, 1676

\bibitem[Carretta \& Gratton\/(1997)]{carre97} Carretta E., Gratton R.G.
 1997, \aap 121, 95


\bibitem[Di Fabrizio et al.\/(2005)]{dif05} Di Fabrizio L.,
Bragaglia A., Tosi M., Marconi G., 2005, \mnras~ in press

\bibitem[Friel et al.\/(2002)]{fri02} Friel E.D., Janes K.A., Tavarez
M., Jennifer S., Katsanis R., Lotz J., Hong L., Miller N. 2002,
\aj~ 124, 2693

\bibitem[Friel et al.\/(2003)]{fri03} Friel E.D., Jacobson H.R.,
Barrett E., Fullton L., Balachandran A.C., Pilachowski C.A. 2003,
\aj~ 126, 2372

\bibitem[Frinchaboy et al.\/(2005)]{fri05} Frinchaboy P.M., Munoz R.R., Majewski S.R.,
Frield E.D., Phelps R.L., Kunkel W.B., 2005, {\tt astro-ph/0411127}


\bibitem[Girardi et al.\/(2000)]{gir00} Girardi L., Bressan A., Bertelli G.,
Chiosi C. 2000, \aaps~ 141, 371

\bibitem[Gratton et al.\/(1996)]{gra96}Gratton R.G., Carretta E., Castelli F.
1996, \aap~ 314, 191

\bibitem[Guarnieri \& Carraro \/(1997)]{gua97} Guarnieri D., Carraro G. 1997,
\aaps~ 121, 451

\bibitem[Kaluzny\/(1994)]{kal94} Kaluzny J. 1994, \aaps~ 108, 151

\bibitem[Kurucz\/(1992)]{kur92} Kurucz R.L. 1992, in IAU Symposium 149,
The Stellar Populations of Galaxies, ed. B. Barbuy \&
A. Renzini (Dordrecht:Kluwer), 225

\bibitem[Phelps \& Janes \/(1996)]{phe96} Phelps R.L., Janes K.A. 1996, \aj~ 111, 1604

\bibitem[Straizys et al. \/(1981)]{str81} Straizys V., Kuriliene G. 1981, Ap\&SS 80, 353

\bibitem[Udry et al. \/(1999)]{Udr99} Udry S., Mayor M., Queloz D. 1999, ASPC 185, 367

\bibitem[Vogt et al.\/(1994)]{vog94} Vogt S.S. et al. 1994,
SPIE 2198, 362

\end{thebibliography}
\end{document}